\documentclass[usenatbib]{mn2e}
\usepackage{graphicx}
\usepackage{times}

\title[Irradiated donor in 4U 1636-536/V801 Ara and 4U 1735-444/V926 
Sco]{Detection of the Irradiated Donor in the LMXBs 4U 1636-536 
(=V801 Ara) and 4U 1735-444 (=V926 Sco)}

\author[J. Casares et al.]
{J. Casares$^1$\thanks{E-mail: jcv@iac.es (JC); cornelis@astro.soton.ac.uk 
(RC); dsteeghs@head.cfa.harvard.edu (DS); pac@saao.ac.za (PAC); rih@phys.lsu.edu
(RIH); kobrien@eso.org (KOB); stroh@clarence.gsfc.nasa.gov (TES)}, 
R. Cornelisse$^{1,2\star}$, D. Steeghs$^{3\star}$, P.A. Charles$^{2,4\star}$, 
R.I. Hynes$^{5\star}$, K. O'Brien$^{6\star}$, 
\newauthor T. E. Strohmayer$^{7\star}$\\
$^1$ Instituto de Astrof\'{\i}sica de Canarias, 38200 La Laguna, Tenerife,
Spain\\
$^2$ School of Physics \& Astronomy, University of Southampton, Southampton, 
SO17 1BJ, UK\\
$^3$ Harvard-Smithsonian Center for Astrophysics, Cambridge, MA 02138, USA\\
$^4$ South African Astronomical Observatory, P.O. Box 9. Observatory 7935, South
Africa\\
$^5$ Department of Physics and Astronomy, 202 Nicholson Hall, Lousiana State,
Baton Rouge, L.A. 70803, USA\\
$^6$ European Southern Observatory, Casilla 19001, Santiago 19, Chile\\
$^7$ Laboratory for High Energy Astrophysics, NASA Goddard Space Flight Center,
Greenbelt, MD 20771, USA}
\begin{document}

\maketitle

\begin{abstract}
Phase-resolved VLT spectroscopy of the bursting Low Mass X-ray 
Binaries 4U 1636-536/V801 Ara and 4U 1735-444/V926 Sco is presented. Doppler 
images of the NIII $\lambda$4640 Bowen transition reveal compact spots which we 
attribute to fluorescent emission from the donor star and enable us to 
define a new set of {\it spectroscopic} ephemerides. 
We measure $K_{\rm em}=277\pm22$ km s$^{-1}$ and $K_{\rm em}=226\pm22$ km 
s$^{-1}$ from the NIII spots in V801 Ara and V926 Sco respectively which 
represent strict lower limits to the radial velocity semi-amplitude of 
the donor stars.
Our new ephemerides provide confirmation that lightcurve maxima 
in V801 Ara and likely V926 Sco occur at superior conjunction of the 
donor star and hence photometric modulation is caused by the visibility 
of the X-ray heated donor.
The velocities of HeII $\lambda$4686 and the broad Bowen blend are 
strongly modulated with the orbital period, with phasing supporting emission 
dominated by the disc bulge. 
In addition, a reanalysis of burst 
oscillations in V801 Ara, using our spectroscopic $T_0$, leads to 
$K_1=90-113$ km s$^{-1}$.
We also estimate the $K-corrections$ for all possible 
disc flaring angles and present the first dynamical constraints on the masses 
of these X-ray bursters. These are $K_2=360\pm74$ km s$^{-1}$, 
$f(M)=0.76\pm0.47$ M$_{\odot}$ and $q=0.21-0.34$ for V801 Ara and 
$K_2=298\pm83$ km s$^{-1}$, $f(M)=0.53\pm0.44$ M$_{\odot}$ and $q=0.05-0.41$ 
for V926 Sco. 
Disc flaring angles $\alpha \ge 12^{\circ}$ and $q\simeq0.26-0.34$ are
favoured for V801 Ara whereas the lack of $K_1$ constraint for V926 Sco
prevents tight constraints on this system. 
Although both binaries seem to have intermediate inclinations, the 
larger equivalent width of the narrrow NIII line in V801 Ara at phase 0.5 
relative to phase 0 suggests that it has the higher inclination of the two.
\end{abstract}


\section{Introduction} \label{introduction}

Low mass X-ray binaries (LMXBs) are interacting binaries where a low mass 
donor transfers matter onto a neutron star or black hole. 
4U 1636-536 (=V801 Ara) and 4U 1735-444 (=V926 Sco) are among the optically 
brighter members in the class of {\it persistent} LMXBs, characterized by  
$L_{\rm x} \simeq 10^{37-38}$ erg s$^{-1}$ and blue spectra with weak
high-excitation emission lines (mainly HeII $\lambda$4686 and the 
CIII/NIII Bowen blend at $\lambda$4640). They also share similar 
properties: they are both atoll sources (as based on the pattern traced 
in X-ray color-color diagrams, see e.g.\citealt{hasin89}) with frequent 
burst activity and short orbital periods (3.80 and 4.65 hrs respectively) 
revealed through optical photometry (\citealt{corbet86}, \citealt{pedersen81}). 
Their lightcurves display shallow sinusoidal modulations which have been 
interpreted as due to the geometrically varying visibility of the irradiated 
donor star (e.g. \citealt{van88}). Therefore, the photometric maxima 
supposedly define orbital phase 0.5 i.e. inferior conjunction of the 
compact object. Note, however, that this assumption requires confirmation 
because photometric maxima can sometimes be associated with asymmetries in the 
disc structure such as the visibility of the irradiated inner disc bulge 
at phase $\sim 0.3$ (e.g. 4U 1822-371 \citealt{hell89}) or superhump
activity (see \citealt{haswell01}). 

Only a few 
spectroscopic studies have been presented on these two binaries up to now. 
For example, \cite{smale91} report H$_{\alpha}$ spectroscopy of V926 Sco 
showing that  the line core is dominated by emission from the disc bulge or
splash region where the gas stream interacts with the outer disc rim. 
On the other hand, \cite{aug98} (A98 hereafter) present radial velocity curves 
of HeII $\lambda$4686 and the Bowen blend at $\lambda$4640 for both V926 Sco 
and V801 Ara. They conclude that these high excitation lines are also tracing 
the motion of the disc bulge.  

V801 Ara is particularly remarkable since it is one of only 14 bursters where 
``burst oscillations" (i.e. nearly coherent high-frequency pulsations) 
have been detected during several thermonuclear X-ray bursts 
(\citealt{giles02}, G02 hereafter). 
Furthermore, a train of burst oscillations was also discovered in a 
13 min  interval during a ``superburst", showing a frequency drift which has 
been interpreted as the orbital motion of the neutron star (\citealt{stro02},
SM02 hereafter). 
By fitting the frequency evolution with a circular orbit 
SM02 constrain the radial velocity amplitude of the neutron star to 
the range $90<K_1<175$ km s$^{-1}$. These constraints on $K_1$ 
will be readdressed and improved in this paper. 
Further constraints on the system parameters and the component masses, 
however, require dynamical information on the donor
star which, unfortunately, is normally overwhelmed by the optical emission 
from the X-ray irradiated disc.    
                                                                                
A new indirect method to extract dynamical information from donor stars in 
persistent LMXBs was proposed by \cite{stee02} (SC02 hereafter). They detected 
narrow high excitation emission lines from the irradiated surface of the donor 
star in Sco X-1 which led to the first determination of its radial velocity 
curve and mass function. These lines are strongest in the Bowen blend, a 
combination of C\,III and N\,III lines which are powered by photoionization and 
fluorescence (respectively) due to UV photons arising from the hot inner 
disk (\citealt{mc75}). This technique has been amply confirmed by subsequent 
studies of the eclipsing ADC (Accretion Disc Corona) and X-ray pulsar 
4U 1822-371 (\citealt{casa03}), the black hole soft X-ray transient 
GX 339-4 during its 2002 outburst (\citealt{hynes03}) and   
Aql X-1 during its 2004 outburst (\citealt{cornel06}).
 
In this paper we apply this method to the X-ray bursters 
4U 1636-536 (=V801 Ara) and 4U 1735-444 (=V926 Sco) and present the first 
detection of the donor stars in these two LMXBs. The paper is organized as
follows: Section 2 summarizes the observation details and data reduction. 
The average spectra and main emission line parameters are presented in Sect. 
3, with multigaussian deconvolution of the Bowen blend. In Section 4 we analyse 
the radial velocities and orbital variability of the strong Bowen blend and 
HeII $\lambda$4686 emission lines. Estimates of the systemic velocities are 
obtained through the Double Gaussian technique applied to the wings of the 
HeII line. Using these systemic 
velocities we compute Doppler tomograms of HeII $\lambda$4686 and the Bowen 
fluorescence NIII $\lambda$4640 which are presented in Section 5. The NIII maps 
display evidence of irradiated emission from the donor star, which is used to
refine the absolute phasing and systemic velocities. Finally, in Section 
6 we provide an improved determination of $K_1$  for V801 Ara and present our 
constraints on the masses in the two binaries.

\section{Observations and Data Reduction} \label{observations}

V801 Ara and V926 Sco were observed on the nights of 23 and 25 June 2003 using 
the FORS2 Spectrograph attached to the 8.2m Yepun Telescope (UT4) at the 
Observatorio Monte Paranal (ESO). A total of 42 spectra of 600s of V801 
Ara and 102 exposures of 200s of V926 Sco were 
obtained with the R1400V holographic grating, covering a complete 
orbital cycle per night for each target. We used a 0.7 arcsec slit width which 
rendered a wavelength coverage of $\lambda\lambda$4514-5815 at 
70 km s$^{-1}$ (FWHM) resolution, as measured from Gaussian fits to the arc 
lines. The seeing was variable between 0.6" -- 1.2" during 
our run. The flux standard Feige 110 was also observed with the 
same instrumental configuration to correct for the instrumental 
response of the detector.

The images were de-biased and flat-fielded, and the spectra subsequently 
extracted using conventional optimal extraction techniques in order to  
optimize the signal-to-noise ratio of the output spectra (Horne 1986). A 
He+Ne+Hg+Cd 
comparison lamp image was obtained in daytime to provide the wavelength 
calibration scale. This was obtained by a 4th-order polynomial fit to 
19 lines, resulting in a dispersion of 0.64 \AA~pix$^{-1}$ and rms scatter 
$<$ 0.05 \AA. Instrumental flexure was monitored through cross-correlation 
between the sky spectra and was found to be very small, always within 14 km
s$^{-1}$ (0.4 pix.) on each night. These velocity drifts were removed from each 
individual spectrum, and the zero point of the final wavelength scale 
was established from the position of the strong OI $\lambda$5577.338 sky line.  
All the spectra were calibrated in flux using observations of the flux 
standard Feige 110. However, due to light loss caused by our narrow slit and 
variable seeing conditions, our flux calibration is only accurate to $\sim$ 50\%. 

\section{Average Spectra and Emission Line Parameters}

%------------------------------------------------------------------------------
\begin{figure}
\centering
\includegraphics[angle=-90,width=84mm]{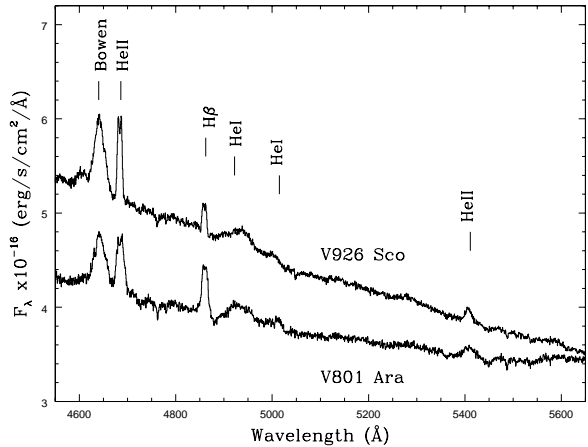}
\caption{Summed spectra of V801 Ara and V926 Sco with the principal emission 
lines indicated.}
\label{fig1}
\end{figure}
%------------------------------------------------------------------------------

Figure~\ref{fig1} presents the average spectra of V801 Ara and V926 Sco in 
$f_{\lambda}$ flux units. They show a blue continuum with broad high excitation 
emission lines of HeII $\lambda$4686, $\lambda$5411, the Bowen blend at 
$\lambda\lambda$4630-50 and H$\beta$, which are typical of X-ray active 
LMXBs. Possible HeI lines at $\lambda$4922 and $\lambda$5015 
are also identified but these are significantly weaker. Table 1 summarizes the 
FWHM, EW and centroid $\lambda_{\rm c}$ of the main emission lines obtained 
through simple Gaussian fits. 
We note that the average spectra (including the line strengths) look very 
similar to those presented by A98, which seems to imply that there has been no 
large, long-term variations between the two data epochs. 
Incidentally, the EWs and FWHMs of all lines do not show any significant 
nigh-to-night variability nor modulation with the orbital period.
Aside from the intrinsically broad Bowen blend, which is 
a blend of at least three CIII/NIII transitions, we note that emission lines 
are a factor $\sim$2 narrower in V926 Sco than in V801 Ara. Given  the 
similarities in their orbital periods, this suggests a projection effect with a 
lower inclination angle for V926 Sco. In addition, the emission line centroids 
in V926 Sco are significantly blueshifted, probably due to a larger 
(approaching) systemic velocity. 
This also applies to the Bowen complex since the difference in
  wavelength between V801 Ara and V926 Sco (-110 km/s) is consistent with 
  the difference in velocities found for the other lines. 
The blue continuum is also steeper for 
V926 Sco, as expected because of its lower reddening (see A98).  

%------------------------------------------------------------------------------
\begin{table}
\centering
\caption[]{Emission line Parameters} 
\begin{tabular}{lcccc}
\hline 
{\em Line} & {\em FWHM} & {\em EW} & {\em Centroid} &  {\em $\Delta V$} \\
 &  {\em (km s$^{-1}$)} & {\em (\AA)} & {\em (\AA)} & {\em (km s$^{-1}$)} \\
\hline
{\bf V801 Ara} & & & \\
Bowen               & 1848 $\pm$ 65 & 4.25 $\pm$ 0.03 & 4643.2 $\pm$ 0.4 & -- \\ 
HeII $\lambda$4686  & 1216 $\pm$ 45 & 3.31 $\pm$ 0.03 & 4685.6 $\pm$ 0.3 & -9  $\pm$ 11 \\
$H\beta$            &  963 $\pm$ 56 & 1.67 $\pm$ 0.02 & 4860.9 $\pm$ 0.3 & -25 $\pm$ 21  \\
\hline
\hline
{\bf V926 Sco} & & & \\
Bowen               & 1662 $\pm$ 65 & 3.99 $\pm$ 0.02 & 4641.5 $\pm$ 0.4 & -- \\ 
HeII $\lambda$4686  &  657 $\pm$ 26 & 2.18 $\pm$ 0.02 & 4683.9 $\pm$ 0.2 & -118 $\pm$ 12 \\
$H\beta$            &  507 $\pm$ 39 & 0.43 $\pm$ 0.02 & 4859.2 $\pm$ 0.4 & -134 $\pm$ 25  \\
\hline
\label{lines}
\end{tabular}
\end{table}
%------------------------------------------------------------------------------

%------------------------------------------------------------------------------
\begin{figure}
\centering
\includegraphics[angle=-90,width=84mm]{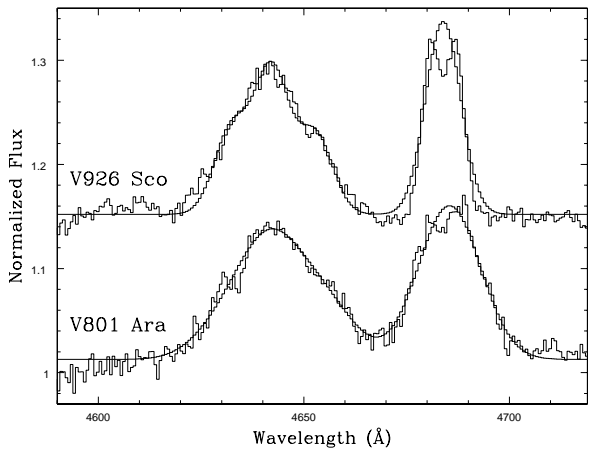}
\caption{Combined fit to HeII $\lambda$4686 and the Bowen blend using three 
Gaussians at $\simeq$4634\AA~ (NIII), $\simeq$4641\AA~(NIII) and 
$\simeq$4651\AA~ (CIII). See text for details.}
\label{fig2}
\end{figure}
%------------------------------------------------------------------------------
 
The Bowen blend mainly consists of emissions corresponding to the NIII 
transitions at $\lambda\lambda$4634,4641,4642 and CIII at  
$\lambda\lambda$4647,4651,4652 (\citealt{scha89}). In an attempt to estimate 
the relative contribution of the different transitions we have performed a 
combined multi-Gaussian fit to 
the average emission profiles of HeII $\lambda$4686 and the Bowen blend for 
the two LMXBs. The fit consists of four Gaussians, one for the HeII line, two 
for the NIII emissions at $\lambda$4634 and $\lambda$4641-2 and another for 
the CIII emissions at $\lambda$4647-52. Free parameters are the 
line widths, which are set to be equal for all the lines, individual line 
centroids and intensities. The line widths are mainly driven by the fit to the
unblended HeII line and hence the latter is effectively used as a template to 
constrain the line profiles within the Bowen blend. Figure~\ref{fig2} presents 
the results of the fit. Because widths are set to be the same for all lines, 
flux ratios are given by simple peak ratios. Our best fit yields NIII ratio 
$I(\lambda4634)/I(\lambda\lambda4641-2)= 0.31 \pm 0.10$ and $0.59 \pm 0.02$ 
for V801 Ara and V926 Sco, respectively. This is a factor $\leq$ 2 lower than 
the theoretical NIII ratio of 0.71, computed by \cite{nuss71} for Bowen 
fluorescence, but consistent with the results of \cite{scha89} for 
Sco X-1. Similarly, Hynes (private communication) finds a NIII ratio in the 
range 0.36-0.56 (depending on time) for GX 339-4 during its 2002 outburst. 
We can also calculate the CIII/NIII ratio 
$I(\lambda\lambda4647-52)/I(\lambda4634 + \lambda\lambda4641-2)$ 
and find $0.38 \pm 0.06$ and $0.35 \pm 0.01$ for V801 Ara and V926 Sco 
respectively. For comparison, we have 
performed the same analysis on average spectra of Sco X-1 and 4U 1822-371, 
using data presented in SC02 and \cite{casa03}, and find 
$0.44 \pm 0.03$ and $0.49 \pm 0.01$ respectively. The lower CIII/NIII ratio in 
V801 Ara and V926 Sco may indicate possible evidence of CNO processed material 
but we have to be cautious here because of the oversimplification in our 
fitting model. In particular, we cannot rule 
out other possible transitions (most likely OII lines at 
$\lambda$4641.8 and $\lambda$4649.1; see \citealt{mc75}) 
contributing to our components at $\sim$4641\AA~and $\sim$4651\AA, 
which so far we have assumed to be dominated by NIII and CIII transitions. 
Furthermore, since the CIII and 
OII lines are not powered by Bowen fluorescence but photoionization, 
different line ratios may stem from differences in efficiency of these two 
excitation mechanisms rather than true CNO abundance variations. 

\section{Radial Velocity Study and Orbital Variability}

\begin{figure}
\centering
\includegraphics[angle=-90,width=84mm]{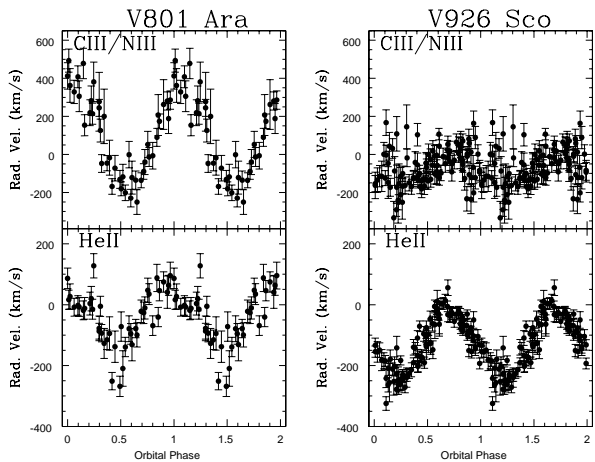}
\caption{Radial velocity curves of the Bowen blend (top) and HeII 
$\lambda$4686 in V801 Ara and V926 Sco.}
\label{fig3}
\end{figure}

Figure~\ref{fig3} displays the radial velocities of the Bowen blend and HeII 
$\lambda$4686 line for V801 Ara and V926 Sco, obtained by cross-correlating 
the individual spectra with Gaussians of fixed FWHM as given in Table 1. 
We have assumed $\lambda_c = 4643.0$ for the central 
wavelength of the Bowen blend. 
The velocities are folded in orbital phase, using the
ephemerides of G02 and A98 for V801 Ara 
and V926 Sco but shifted in phase by +0.5 to make phase zero 
coincide with the inferior conjunction of the donor star. These will be  
called the {\it photometric ephemerides} and will be used throughout this 
Section. The accumulated phase uncertainty at the time of our observations is 
$\pm 0.06$ for both systems. Note that this phase convention assumes that 
lightcurve maxima are driven by the visibility of the irradiated donor star.  

The radial velocity curves for V801 Ara show maxima at phase $\sim$ 1. 
This phasing suggests that the velocity variations contain a significant 
component arising in the disc bulge (which has its maximum visibility at 
phase $\sim$ 0.75), caused by the 
interaction of the gas stream with the outer edge of the disc, as 
is typically seen in persistent LMXBs (e.g. 4U 1822-371; \citealt{cowley82}). 
On the other hand, the radial velocity curves for V926 Sco 
show maxima at phase $\sim$ 0.7. 
By comparing the HeII curves in the two binaries we clearly see  
evidence for a large systemic velocity of $\sim$ -150 km s$^{-1}$ in 
V926 Sco, in good agreement with \cite{smale91}. We also note that the 
CIII/NIII blend in V801 Ara is modulated with 
a much larger amplitude than in V926 Sco i.e. 280 km s$^{-1}$ versus 
70 km s$^{-1}$, respectively. 

%------------------------------------------------------------------------------
%------------------------------------------------------------------------------

\begin{figure}
\centering
\includegraphics[angle=-90,width=84mm]{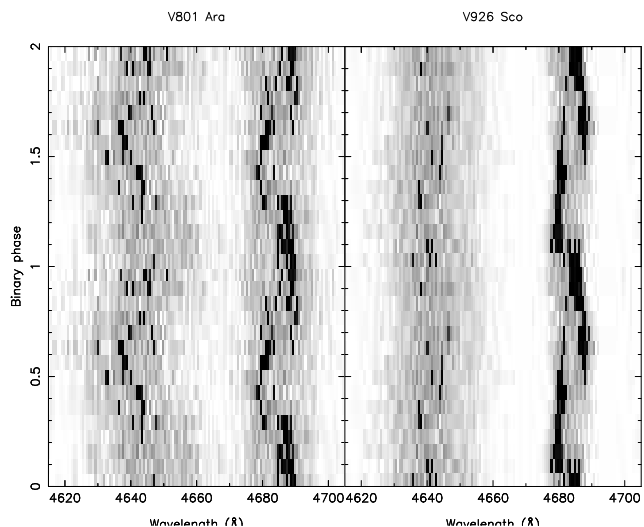}
\caption{Trailed spectra showing the orbital evolution of the Bowen blend 
and HeII $\lambda$4686 in 15 phase bins using the {\it photometric ephemerides}}
\label{fig4}
\end{figure}

Figure~\ref{fig4} presents the trailed spectra of the Bowen blend and 
HeII$\lambda$4686 for V801 Ara and V926 Sco, after co-adding our individual 
spectra in 15 phase bins to improve statistics. The core of the emission lines 
in V801 Ara show clear S-wave components. The S-wave in the Bowen blend is 
quite narrow and it may arise from the heated face of the donor star, as 
observed in Sco X-1 (SC02). 
The S-wave in HeII is rather extended and is likely produced in a 
region with a large velocity dispersion such as the disc bulge. 
On the other hand, the Bowen blend in V926 Sco is rather noisy and does not 
exhibit any clear components that are visible by eye. The HeII trailed 
spectra do show a clear orbital modulation with complex structure. 
 
%------------------------------------------------------------------------------
%------------------------------------------------------------------------------
    
\begin{figure}
\centering
\includegraphics[angle=0,width=84mm]{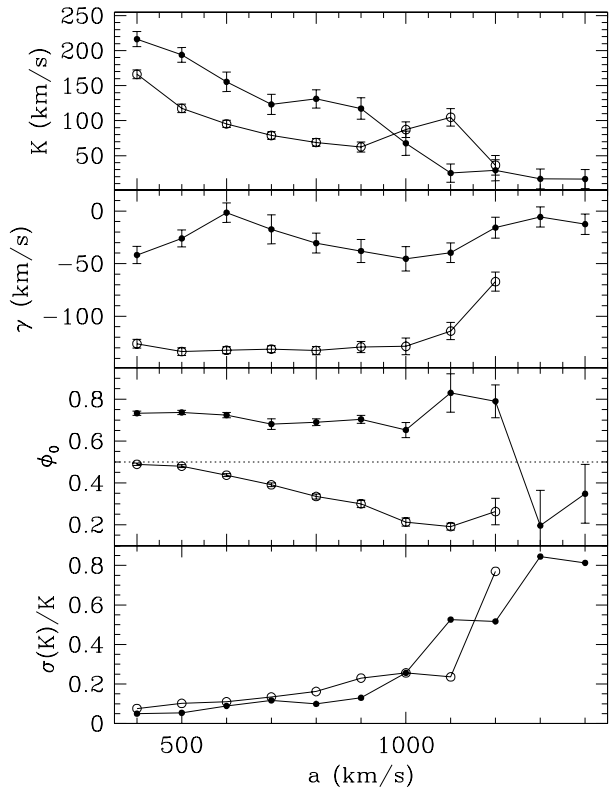}
\caption{Diagnostic Diagram for HeII $\lambda$4686 in V801 Ara (solid circles) 
and V926 Sco (open circles). The dotted horizontal line marks $\phi_{0}=0.5$ 
or the expected inferior conjunction of the compact object, according to the
{\it photometric ephemerides}.}
\label{fig5}
\end{figure}

Following our work on Sco X-1 (SC02) and 4U 1822-371
(\citealt{casa03}), we have applied the 
double-Gaussian technique (\citealt{sch80}) to the wings of the HeII 
$\lambda$4686 line in an attempt to estimate the radial velocity curve of 
the compact object. The trailed spectra presented in Fig.~\ref{fig4} 
demonstrate that the line cores are dominated by 
strong, complex, low-velocity components associated with asymmetric emission 
from the outer disc and/or the donor star, which we want to avoid. 
Therefore, by convolving the emission line with a double-Gaussian filter of 
sufficiently large Gaussian 
 separation we can extract radial velocity curves from the wings of 
the profile, which are expected to follow the motion of the compact star. We 
have used a double-Gaussian bandpass with $FWHM=100$ km s$^{-1}$ and 
relative Gaussian separations in the range  
$a= 400-1400$ km s$^{-1}$ in steps of 100 km s$^{-1}$. In order to 
improve statistics, we have co-added our spectra in 15 phase bins using the 
{\it photometric ephemerides}. The radial velocity curves
obtained for different Gaussian separations are subsequently fitted with 
sine-waves of the form $V(\phi)=\gamma + K \sin 2\pi (\phi-\phi_{0})$, fixing 
the period to the orbital value. The best fitting parameters 
are displayed as a function of the Gaussian separation $a$ in a Diagnostic 
Diagram (see Fig.~\ref{fig5}). 

%------------------------------------------------------------------------------
%------------------------------------------------------------------------------

In both cases we see how the line cores are dominated by high-amplitude 
($K\sim 200$ km s$^{-1}$) S-waves which fade as we move to the line wings,
where lower $K-$amplitudes of a few tens of km s$^{-1}$ are found. On the other
hand, the $\gamma$ velocities are very steady throughout the profiles, with 
average values of~$\sim -30$ (V801 Ara) and~$\sim -130$ km s$^{-1}$ (V926 Sco). 
The blue-to-red crossing phase for V926 Sco 
displays a smooth decreasing trend from the line core to the wings, 
whereas it is rather constant for V801 Ara, with an average value of 
$\phi_{0} \sim 0.7$. 
We estimate that the velocity points start to be
corrupted by continuum noise for Gaussian separations larger than 
$a\sim1000$ as indicated by the diagnostic parameter $\sigma(K)/K$ 
(see \citealt{shafter86}). Therefore, we decided to adopt the average values 
for the parameters obtained from the last two separations before $\sigma(K)/K$ 
starts to rise i.e. $a=900-1000$ km s$^{-1}$ for V801 Ara and 1000-1100 km 
s$^{-1}$ for V926 Sco. 
This yields $K_1 = 93 \pm 25$ km s$^{-1}$, $\gamma = -42 \pm 4$ km 
s$^{-1}$,  $\phi_{0}=0.68 \pm 0.03$ for V801 Ara and $K_1 = 96 \pm 9$ 
km s$^{-1}$, $\gamma = -121 \pm 7$ km s$^{-1}$, $\phi_{0}=0.20 
\pm 0.01$ for V926 Sco. The errorbars have been adjusted to incorporate the
scatter between different values within our preferred range. Note that 
$\phi_{0}$ of V801 Ara is delayed by 
$\sim0.18$ with respect to the inferior conjunction of the compact 
object (dotted line in Fig.~\ref{fig5}), as predicted by the {\it photometric 
ephemerides}. 
This is classically observed in interacting binaries and interpreted as 
contamination of the line wings by residual emission from the 
disc-bulge/hot-spot (\citealt{marsh98}).   
On the other hand, $\phi_{0}$ of V926 Sco leads the inferior 
conjunction of the compact star by $\sim0.30$, according to the 
{\it photometric ephemerides}. Such a large 
shift is unexpected and a full discussion is diverted to Sect. 6.
Furthermore, we also note the very asymmetric distributions obtained 
in the HeII Doppler maps (see next Sect.) which may invalidate the 
double-Gaussian technique. Therefore, the constraints on zero-phase, 
$\gamma$-velocity and $K_1$ (for V801 Ara) will be superseded in the 
following sections. 

\section{Bowen Fluorescence from the Irradiated Companion}   

Doppler tomography enables us to map the brightness 
distribution of a binary system in velocity space through combining 
orbital phase spectra (see details in \citealt{marsh01}). This is 
particularly effective when dealing with weak emission
features which are barely detected or embedded by noise in individual 
spectra, such as our narrow Bowen emission lines. In order to compute Doppler 
images of the NIII $\lambda$4640 fluorescence line in V801 Ara and V926 Sco 
we have rectified the individual spectra by subtracting a low order 
spline fit to the continuum regions and rebinned them into a uniform 
velocity scale of 37 km s$^{-1}$ per pixel. Doppler images were subsequently 
computed using the {\it photometric ephemerides} and $\gamma=-42$ and $-121$ 
km s$^{-1}$ for V801 Ara and V926 Sco, respectively. The maps show compact 
spots shifted in phase by $+0.03$ and $-0.18$ with respect to the expected 
location of the donor star in  the velocity maps (i.e. along the vertical
$V_{\rm y}$ axis)\footnote{Note that the greyscale in the NIII maps have 
been set to enhance the contrast of the sharp components. This is the reason 
why these maps do not show any trace of emission from the underlying broad 
component.}. Assuming that these spots 
are produced on the irradiated hemisphere of the donor star, as has been
shown to be the case in Sco X-1 (SC02) and 4U 1822-371 
(\citealt{casa03}), we correct the previous {\it photometric ephemerides} and 
derive the following {\it spectroscopic ephemerides}, which will be used in the 
remainder of the paper:   

\begin{equation} 
T_{0}(HJD) = 245 2813.531(2) + 0.15804693(16) E    
\end{equation}

\begin{equation} 
T_{0}(HJD) = 245 2813.495(3) + 0.19383351(32) E      
\end{equation}

\noindent
where the zero-phase error comes from the uncertainty in the centroid
position of the spots. Equation (1) corresponds to the ephemerides of  
V801 Ara and equation (2) to V926 Sco.
With these ephemerides, the HeII $\lambda$4686 maps of the two binaries show 
a crescent shape brightness distribution pointing towards emission from an 
extended disc bulge.  

%------------------------------------------------------------------------------
\begin{figure}
\centering
\includegraphics[angle=-90,width=84mm]{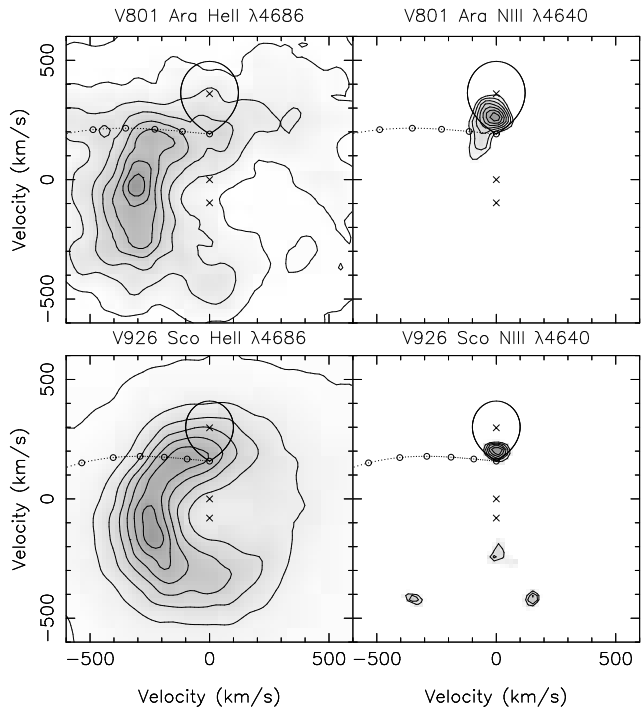}
\caption{Doppler maps of HeII $\lambda$4686 (left panels) and NIII 
$\lambda$4640 (right panels) for V801 Ara and V926 Sco, computed using our 
{\it spectroscopic ephemerides}. Systemic velocities 
$\gamma = -34$ km s$^{-1}$ and $\gamma = -140$ km s$^{-1}$ were adopted for 
V801 Ara and V926 Sco, respectively. For comparison we also plot the gas
stream trajectory (in units of 0.1 R$_L1$) and Roche lobe of the donor star 
for the case of $K_2=360$ km s$^{-1}$ (V801 Ara) and $K_2=298$ km 
s$^{-1}$  (V926 Sco). A mass ratio $q=0.27$ has been assumed for the two
binaries. 
 }
\label{fig6}
\end{figure}
%------------------------------------------------------------------------------

\begin{figure}
\centering
\includegraphics[angle=-90,width=84mm]{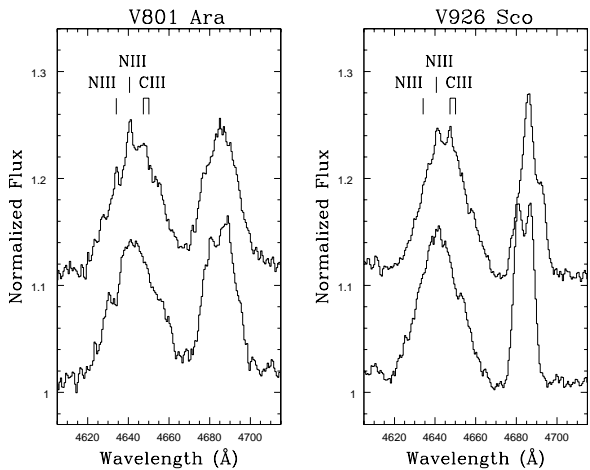}
\caption{Average spectra of V801 Ara (left panel) and V926 Sco (right panel).
Top spectra have been Doppler corrected in the rest frame of the donor using 
the spectroscopic ephemerides and the $K$-velocity amplitudes derived from the 
bright spots in the NIII $\lambda4640$ Doppler images. The narrow NIII and 
CIII lines are clearly seen in the Bowen blend. These are smeared out and cannot
be detected in the straight averages shown at bottom. Noise has been filtered 
using a 2 pixel boxcar smoothing.}
\label{fig7}
\end{figure}

Following SC02 we have used the HeII Doppler maps to refine the 
systemic velocities derived in the previous section. The 
$\chi^2$ value of the map was calculated for a range of $\gamma$'s, and the 
best fit in terms of mimimal $\chi^2$ was achieved for $\gamma\simeq -39 $  
km s$^{-1}$ (V801 Ara) and $\gamma \simeq -132 $  km s$^{-1}$ (V926 Sco). 
The use of an incorrect systemic velocity has the effect  
of blurring bright spots into , as opposed to V801 Ara,elongated ``defocused" 
features in Doppler 
images. Therefore, as a further test, we have also computed NIII  
$\lambda$4640 maps for a set of $\gamma$-velocities in the range 
-200 -- +200 km s$^{-1}$ and we looked for the best focused NIII spots by 
computing the
skewness using box sizes of 5, 10 and 20 pixels. We find that the most 
symmetric and compact spots are found for $\gamma$ values in the
range -29 -- -39 km s$^{-1}$ (V801 Ara) and 137-143  km s$^{-1}$ (V926 Sco). 
Therefore, we decided to adopt $\gamma=-34 \pm 5$ km s$^{-1}$ (V801Ara) 
and $\gamma=-140 \pm 3$ km s$^{-1}$ (V926 Sco) and these are the 
values that we have used in the final Doppler maps, which are presented in
Fig.~\ref{fig6}. The NIII maps reveal compact bright spots 
which we interpret as irradiated emission from the inner hemisphere of
the donor star. These spots are obviously located along the vertical axis
because of our ephemerides definition. We have calculated the position of 
these spots by computing their centroids using the centering algorithm 
CENTROID in IRAF\footnote{IRAF is distributed by the National Optical 
Astronomy Observatories, which are operated by the Association of Universities 
of Research in Astronomy, Inc., under cooperative agreement with the
Nacional Science Foundation.} 
and find $K_{\rm em}= 277 \pm 22$ km 
s$^{-1}$ (V801 Ara) and $226 \pm 22$ km s$^{-1}$ (V926 Sco). We note that 
$K_{\rm em}$ is very weakly dependent on $\gamma$, with only a 10 km s$^{-1}$ 
drift when $\gamma$ varies by $\pm 20$ km s$^{-1}$ around our central value. 
Therefore, our $K_{\rm em}$ determinations seem very robust and they represent 
strict lower limits to the true radial velocity semiamplitude $K_2$ of the 
donor stars because they must arise from the irradiated hemisphere facing the 
neutron star. Further confirmation of our system parameters is provided by 
the appearance of the sharp CIII/NIII transitions after coadding all the 
spectra in the rest frame of the NIII $\lambda$4640 emission line region 
(see Fig.~\ref{fig7}). 
Note that we have not attempted to compute Doppler images of the H$\beta$ 
lines because these are contaminated by phase-variable absorption components, 
as was also the case in 4U 1822-371 (e.g. \citealt{casa03}). Absorption 
violates the principles of Doppler Tomography and makes it very difficult to
interpret the corresponding Doppler images.     

%------------------------------------------------------------------------------

%------------------------------------------------------------------------------

\section{DISCUSSION}

The detection of NIII $\lambda$4640 fluorescence emission from the irradiated 
donor in V801 Ara and V926 Sco provides new absolute (spectroscopic) 
ephemerides and opens the possibility to derive the first constraints on the 
dynamical masses of these two LMXBs. In the light of the new ephemerides, the
photometric lightcurve maxima of V801 Ara take place at phase $0.47 
\pm 0.06$ (with the uncertainty dominated by error propagation of the G02 
ephemerides) and, therefore, they are consistent with X-ray irradiation of the 
donor star. On the other hand, lightcurve maxima in V926 Sco are located at
orbital phase $0.82 \pm 0.06$. This is in between the maximum visibility of 
the irradiated donor and the disc-bulge and hence it would imply that 
lightcurve maxima arise by a combination of these two emitting regions. 
However, we note that the uncertainty in the photometric ephemerides of V926 
Sco is likely to be higher than stated in A98 due to the limited number of
arrival times fitted and their large scatter (Augusteijn, private
communication). Therefore, it is possible that lightcurve maxima in V926 Sco are
also consistent with superior conjunction of the irradiated donor star. This is
the most likely scenario because it would be hard to understand an emission
source peaking at phase $\sim 0.8$ in an irradiation dominated environment
without invoking some strange (and stable) disc configuration.  

We have also determined the systemic velocities for the two binaries i.e. 
$-34\pm5$ km s$^{-1}$ (V801 Ara) and $-140\pm3$ km s$^{-1}$ (V926 Sco). 
These can be compared with $V_{\rm r}$, the radial velocity,  
relative to the Local Standard of Rest, due to Galactic 
rotation at the location of the binaries. For the case of 
V801 Ara, we take $l=332.9^{\circ}$, $b=-4.8^{\circ}$ 
and $d=6-7$ kpc, based on peak fluxes of radius-expansion 
X-ray bursts for $M_1=1.4-2$ M$_{\odot}$ (\citealt{galloway05}). Assuming 
the Galactic rotation curve of \cite{naka03} we find 
$V_{\rm r}$ in the range -103 to -111 km s$^{-1}$. This is much larger 
than our systemic velocity which could be explained through the recoil 
velocity gained by the binary during the supernova explosion which formed 
the neutron star. Thus, we can set a lower limit to the kick velocity in 
V801 Ara of $\simeq70$ km s$^{-1}$. Regarding V926 Sco, we take 
$l=346.1^{\circ}$, $b=-7.0^{\circ}$ and a distance $d=9.1$ kpc 
(for a canonical 1.4 M$_{\odot}$ neutron 
star, see A98). The Galactic rotation curve at the position of 
V926 Sco yields $V_{\rm r}= -111$ km s$^{-1}$, close but 
significantly lower than our systemic velocity. The difference might also be 
ascribed to a kick velocity received by the neutron star at birth or,
alternatively, a neutron star mass of $\simeq$1.7 M$_{\odot}$. Note that the 
latter implies a larger distance of $d=10$ kpc and hence $V_{\rm r}=-140$ km 
s$^{-1}$.  

%------------------------------------------------------------------------------

\begin{figure}
\centering
\includegraphics[angle=0,width=84mm]{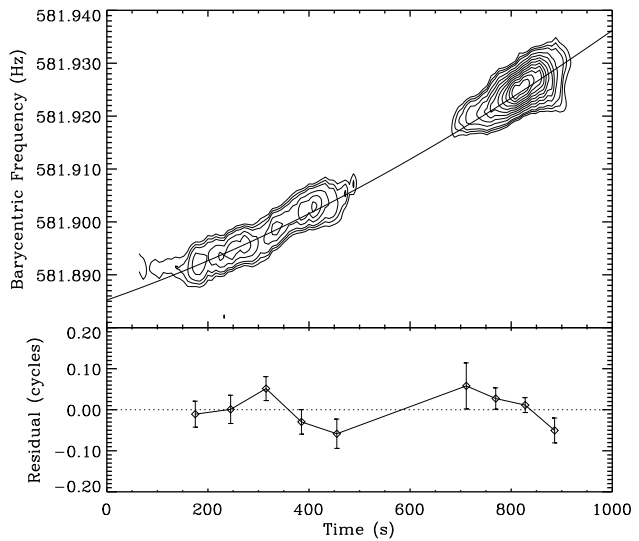} 
\caption{Top panel: Dynamic Fourier power contours showing the excursion 
in oscillation frequency during the 2001 February 22 superburst from 4U 
1636-53 (V801 Ara) with the best fitting orbit model overlayed. Time zero for 
the x-axi is MJD 51962.7102530038 (TDB) 
Bottom panel: Pulse phase residuals for the best fitting model.} 
\label{fig8}
\end{figure}
%------------------------------------------------------------------------------

\subsection{System Parameters for V801 Ara}   

The NIII $\lambda$4640 spot in the Doppler map yields $K_{\rm em}= 277 \pm 
22$ km s$^{-1}$. It must arise on the inner hemisphere of the donor
and hence it sets a lower limit to the velocity semi-amplitude of the 
companion's center of mass $K_2$. In order to find the real $K_2$ we need to
calculate the $K-correction$ or $K_{\rm em}/K_2$ for the case of emission 
lines in illuminated atmospheres, which depends mainly on the binary mass 
ratio $q=M_2/M_1$ (with $M_2$, $M_1$ the masses of the donor and compact 
star respectively) and the disc flaring angle $\alpha$. The $K-correction$ has 
been calculated by \cite*{munoz05} using an irradiation binary code which 
includes shadowing by an axisymmetric flared disc. The results are 
tabulated as a function of $q$, $\alpha$ and the inclination angle $i$ in 
Table 1 of their paper, although the dependence on $i$ is very weak. The 
$K-correction$ is constrained between $\alpha=0$ (i.e. we neglect disc 
shadowing) and the geometric limit set by emission from the irradiated limb of 
the donor, i.e. $K_{\rm em}/K_2 < 1-0.213 q^{2/3} (1+q)^{1/3}$ (see 
\citealt{munoz05}). 

An upper limit on $q$ is established by taking $M_1\ge1.4 $M$_{\odot}$ and 
$M_2\le 0.48$ M$_{\odot}$, the largest possible zero-age main sequence (ZAMS) 
star fitting in a 3.79 hr period Roche lobe (\citealt{tout96}), which leads to 
$q\le 0.34$. As a first approach, one can assume the empirical Mass-Radius 
relation for low-mass stars in cataclysmic variables and LMXBs (see e.g. 
\citealt{smith98}, \citealt{warner95}) which yields $M_2 \simeq 0.32$. This, 
combined with a canonical neutron star of $M_1\simeq1.4$ M$_{\odot}$, 
would lead to 
a plausible $q\sim 0.23$. The $K-correction$ for $K_{\rm em}=277\pm22$ km 
s$^{-1}$ and $q\sim0.23$ would yield $303<K_2<404$ km s$^{-1}$, where we have 
adopted the coefficients for the case $i=40^{\circ}$ in Table 1 of 
\cite{munoz05} because V801 Ara is not eclipsing.
  
A more refined $K-correction$ requires a knowledge of $q$. This can be
constrained by the rotational broadening $V_{\rm rot} \sin i$ of the companion
star which, for the case of synchronous rotation, is related to $q$ through 

\begin{equation}
V_{\rm rot} \sin i \simeq 0.462 K_2 q^{1/3} (1+q)^{2/3}
\end{equation}

\noindent
(\citealt{wadehorne88}). 
A lower limit to $V_{\rm rot} \sin i$ can be estimated from the width 
of the sharp NIII fluorescence emission in the Doppler corrected average 
spectrum. This is because emission lines arise not from the entire
Roche lobe but from the irradiated part only. 
A multigaussian fit to the Bowen profile presented in Fig.~\ref{fig7} gives 
$FWHM (NIII \lambda4640) = 140 \pm 21$ km s$^{-1}$,  which includes the effect 
of our intrinsic instrumental resolution (70 km s$^{-1}$). This has been 
accounted for by broadening a Gaussian template of $FWHM=70$ km s$^{-1}$ 
between 10 and 200 km s$^{-1}$, in steps of 10 km s$^{-1}$, using a Gray 
rotational profile (\citealt{gray92}) without limb-darkening, because 
fluorescence lines arise in optically thin conditions. Our simulation 
indicates that $V_{\rm rot} \sin i=92\pm16$ km s$^{-1}$ 
is equivalent to $FWHM = 140 \pm 21$ km s$^{-1}$. On the other hand, by 
substituting $V_{\rm rot} \sin i \ge 76$ km s$^{-1}$ and the $K-correction$ for 
$\alpha>0^{\circ}$ into equation (3), one finds a secure lower limit $q\ge 
0.08$. 

An upper limit to the rotational broadening of the donor star is set by 
$V_{\rm rot} \sin i=2\pi R_2 \sin i /P$. By assuming that the donor 
must be more evolved than a ZAMS star within a 3.79 hr Roche lobe, then 
$R_2\le 0.44 R_{\odot}$ (\citealt{tout96}). This, together with  
$i\le 78^{\circ}$ (lack of X-ray eclipses for $q\ge0.08$), yields 
$V_{\rm rot} \sin i \le 138$ km s$^{-1}$. 

%------------------------------------------------------------------------------

\begin{figure}
\centering
\includegraphics[angle=-90,width=84mm]{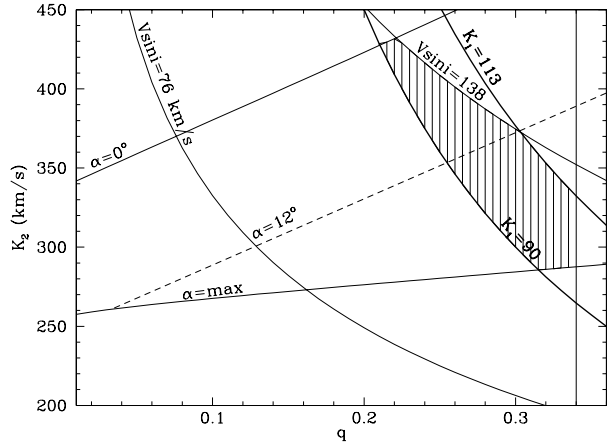} 
\caption{Constraints on $K_2$ and $q$ for V801 Ara. The disc flaring angle 
must be between $\alpha=0^{\circ}$ and $\alpha=max$, whereas the rotational 
broadening is constrained between the observed $V_{\rm rot} \sin i \simeq 76$ 
km s$^{-1}$ and 138 km s$^{-1}$, the maximum allowed for a ZAMS star in a 3.8hr 
period and for $i\le 78^{\circ}$. 
The shaded area represents the $K_1$ solution obtained from a reanalysis of 
the superburst pulsations. These constrain the binary mass ratio to be 
$q=0.21-0.34$ and $K_2=286-433$ km s$^{-1}$. For comparison we also mark  
(dashed line) the solution for a hypothetic disc flaring angle of 
$\alpha=12^{\circ}$.}
\label{fig9}
\end{figure}
%------------------------------------------------------------------------------

Further constraints are provided by the study of burst oscillations which 
can set limits to the neutron star's projected velocity $K_1 (= q K_2)$.  
SM02 derived $90\le K_1\le 175$ km s$^{-1}$ by fitting the frequency
drift of highly coherent X-ray pulsations observed in an 800s interval
during a superburst. They used a circular orbit model and fixed the zero phase 
to the ephemeris of G02. 
We have used our new spectroscopic ephemeris to reanalyse the superburst
pulsation data in order to better constrain $K_1$.  Based on our spectroscopic
ephemeris, the pulsation interval during the superburst from V801 Ara 
comes slightly earlier in orbital phase by 0.032 cycles as compared to the
G02 ephemeris. We fit the pulsation data (see SM02 for details on the phase 
timing analysis) to a circular orbit model with the
reference epoch fixed to our new $T_0$, and we also fixed the orbital period
(using the G02 value). This
leaves two free parameters, the projected neutron
star velocity, $K_1$, and the rest-frame spin frequency, $\nu_0$.  We find
acceptable fits  with a reference epoch within the $\pm 1 \sigma$ range for
our spectroscopic $T_0$.  The inferred $K_1$ velocity ranges from 90 - 113 km 
s$^{-1}$ as $T_0$ ranges over $\pm 1\sigma$.  
The best fit $\chi^2$ values range from 14.6 to 10.6 (with 7 dof) over this 
same range, that is, higher velocities are modestly favored, but given the 
shortness of the pulse
train compared to the orbital period, we do not consider these differences 
as significant. With the reference epoch fixed, the statistical error on the
velocity is much smaller than the range given above, so 90 - 113 km s$^{-1}$ is 
a robust range for $K_1$ at the $\pm 1\sigma$ limits of $T_0$.  The best fit 
$\nu_0$ ranges from 582.04143 to 582.09548 Hz, and which 
may help constrain future searches for a persistent millisecond pulsar in V801  
Ara. Figure ~\ref{fig8} summarizes the results of the new timing analysis. 

Fig.~\ref{fig9} summarizes all our restrictions in the $K_2-q$ parameter 
space i.e. $0^{\circ}\le \alpha\le max$, $V_{\rm rot} \sin i =76-138$ km 
s$^{-1}$ and $K_1=90-113$ km s$^{-1}$. The shaded area 
indicates the region allowed by our constraints, which yields $q=0.21-0.34$ 
and $K_2=286-433$ km s$^{-1}$. These imply a mass function $f(M)=M_1 
\sin^3 i /(1+q)^2=0.76\pm0.47$ M$_{\odot}$ and hence $M_1 
\sin^3 i=1.24\pm0.77$ M$_{\odot}$. 
Clearly the large uncertainties are dominated by the error in 
$K_{\rm em}$ and the uncertainty in the disc flaring angle.

\subsection{System Parameters for V926 Sco}   

The same reasoning can be applied to V926 Sco. The spot in the NIII 
$\lambda$4640 map yields $K_{\rm em}=226 \pm 22$ km s$^{-1}$ whereas 
the ZAMS Mass-Radius relation for a 4.65 hr Roche lobe leads to $M_2 
\le 0.58 M_{\odot}$ (\citealt{tout96}) and hence $q\le 0.41$.
On the other hand, we measure $FWHM (NIII \lambda4640) = 115 \pm 23$ km 
s$^{-1}$ which, after deconvolution of the instrumental resolution using Gray
rotation profiles as above, yields $V_{\rm rot} \sin i \ge 71 \pm 21$ km 
s$^{-1}$. This lower limit, combined with the $K-correction$ for 
$\alpha>0^{\circ}$ and equation (3), yields $q\ge 0.05$ and, hence 
$i\le 80^{\circ}$. And, under the assumption that the donor star is more 
evolved than a ZAMS star, the Roche lobe geometry implies $R_2 \le 0.54 
R_{\odot}$ which, for $i\le 80^{\circ}$, leads to $V_{\rm rot} \sin i 
\le 137$ km s$^{-1}$. 

\begin{figure}
\centering
\includegraphics[angle=-90,width=84mm]{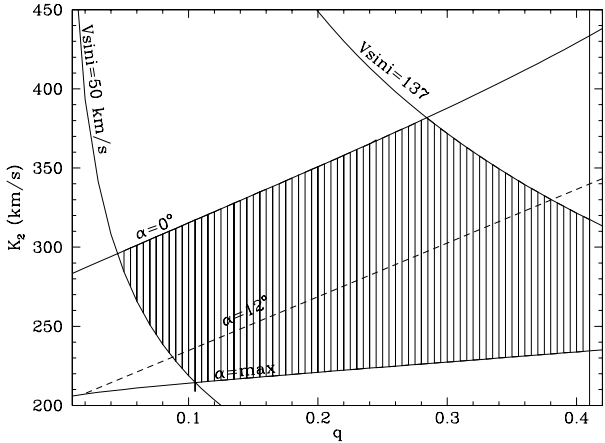} 
\caption{Same as Fig.~\ref{fig9} but for V926 Sco, but with no pulsation 
constraints yet.}
\label{fig10}
\end{figure}

All these restrictions translate into constraints on the $K_2-q$ plane 
which are presented in Fig.~\ref{fig10}. The allowed region results in 
$q=0.05-0.41$ and $K_2=215-381$ km s$^{-1}$, depending on the value of 
$\alpha$. These numbers imply  $f(M)=0.53\pm0.44$ 
M$_{\odot}$ and, thus, $M_1 \sin^3 i= 0.80 \pm 0.71$ M$_{\odot}$.  
Unfortunately we do not have any contraints on $K_1$ from  
burst oscillations and hence our mass restrictions are not as 
well constrained as for V801 Ara.  
Our best estimates of the system parameters for both objects 
are presented in Table 2.

%------------------------------------------------------------------------------

%------------------------------------------------------------------------------

%------------------------------------------------------------------------------
\begin{table}
\centering
\caption[]{System Parameters.~$T_0$ indicates zero phase or inferior conjunction
of the donor star. Orbital periods $P_{\rm orb}$ are from G02 and A98.} 
\begin{tabular}{lcc}
\hline 
Parameter  & {\em 4U 1636-536} & {\em 4U 1735-444} \\
           & {\em V801 Ara}  & {\em V926 Sco}  \\
\hline
$P_{\rm orb}$ (days)       & 0.15804693(16) & 0.19383351(32)         \\
$T_0$ (HJD$-$2\,452\,000)  & 813.531 $\pm$ 0.002    & 813.495 $\pm$ 0.003 \\
$\gamma$ (km~s$^{-1}$)     & -34     $\pm$ 5        & -140    $\pm$ 3     \\
$K_{\rm em}$ (km~s$^{-1}$) & 277     $\pm$ 22       & 226     $\pm$ 22    \\
$q$ ($M_2/M_1$)            &      0.21-0.34         &       0.05-0.41     \\
$K_1$ (km~s$^{-1}$)        & 90-113         &     ---    \\
$K_2$ (km~s$^{-1}$)        & 360     $\pm$ 74       & 298     $\pm$ 83    \\
$f(M)$ (M$_\odot$)         & 0.76    $\pm$ 0.47     & 0.53    $\pm$ 0.44  \\
$i$ (deg)                  & 36-60                  & 27-60               \\
\hline
\label{param}
\end{tabular}
\end{table}
%------------------------------------------------------------------------------
 
\subsection{Constraints on the inclination}

Further constraints on the stellar masses requires a knowledge of the 
binary inclination. While strict upper limits are set by the absence 
of X-ray eclipses, lower limits can be established by combining 
$M_1 \le 3.1$ M$_{\odot}$ (the maximum mass allowed for a stable 
neutron star, \citealt{rho74}) with our mass function and $q$ restrictions. 
This leads to $i=36^{\circ}-74^{\circ}$ and $i=27^{\circ}-80^{\circ}$ for 
V801 Ara and V926 Sco, respectively. In addition, the physical model of 
\cite{frank87} indicates $i\le60^{\circ}$ due to the lack of X-ray dips. 

\begin{figure}
\centering
\includegraphics[angle=-90,width=84mm]{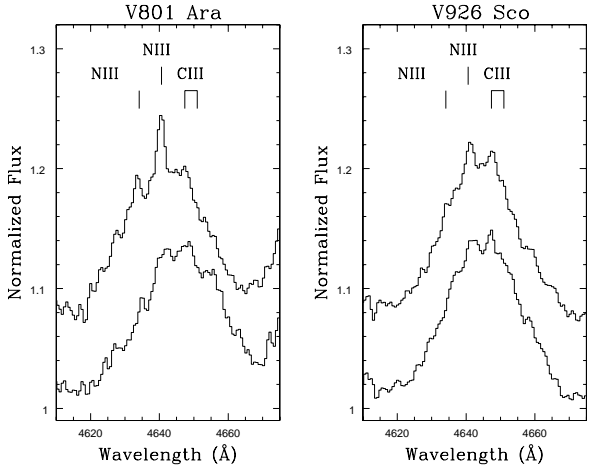}
\caption{Doppler corrected averages of V801 Ara (left panel) and V926 Sco 
(right panel) in the rest frame of the donor star. Top spectra present
averages between orbital phases 0.25-0.75 and bottom spectra between 
0.75-0.25. The sharp CIII/NIII components in V801 Ara are clearly more  
intense around phase 0.5 than 0, suggesting a higher inclination
binary. Spectra have been smoothed by a 2 pixel boxcar.}
\label{fig11}
\end{figure}

On the other hand, we mentioned in Section 3 that the factor $\sim$2 narrowness 
of the HeII $\lambda$4686 profile in V926 Sco with respect to V801 Ara 
indicates a lower inclination. To test this hypothesis
we have produced Doppler corrected averages for V801 
Ara and V926 Sco by coadding all the spectra in two bins centered at orbital 
phases 0.0 and 0.5, when the visibility of the irradiated face of the donor
is minimum and maximum, respectively. These spectra are presented in 
Fig.~\ref{fig11} and show a marked difference for the two binaries: the narrow 
CIII/NIII lines become significantly enhanced around phase 0.5 (top spectra) 
for V801 Ara, but not much difference is seen in V926 Sco between phase 0.5 and
phase 0. This clearly supports a higher inclination angle in V801 Ara. 
Furthermore, the relative contibution of the NIII $\lambda$4640 line with 
respect to the broad base, as estimated through a multigaussian fit, is a 
factor $\sim 2$ larger for V801 Ara than for V926 Sco. This can be taken 
as the relative contributions of the heated donor and disc, and hence, 
strongly suggests that V801 Ara is seen at a higher inclination angle than 
V926 Sco. This seems at odds with the fact that the optical lightcurves 
in V801 Ara and V926 Sco display similar amplitudes, $A\simeq$ 0.2 mags  
(\citealt{vanamer87}, \citealt{van90}). However, \cite{dejong96} have shown 
that lightcurve amplitudes are more sensitive to $\alpha$ than $i$, which  
might simply imply a thicker disc in V926 Sco  and 
hence lightcurve amplitudes cannot be taken as a simple indication of the 
inclination angle. 

\subsection{Mass Estimate and Evolutionary Scenarios} 

\cite{meyer82} analyzed the
effects of X-ray heating in the vertical structure of discs and demonstrated 
that these are strongly correlated. Even in the absence of irradiation, they
find a disc opening angle of $\sim6^{\circ}$. De Jong et al. (1996) 
modelled the effects of irradiation in optical lightcurves and, by
comparing with observations of LMXBS, they infer a mean disc opening angle of 
12$^{\circ}$. Most of their systems are short period LMXBs, just like  
V801 Ara and V926 Sco, and hence it seems justified to speculate what our  
system parameters would be in this particular case. 

Regarding V801 Ara, $\alpha=12^{\circ}$ leads to  
$q\simeq0.25-0.30$, $K_2\simeq356-375$ km s$^{-1}$ (see dashed line 
in Fig.~\ref{fig9}) and hence $f(M)\simeq 0.81 \pm 0.07$ M$_{\odot}$ and 
$M_1 \sin^3 i \simeq 1.32 \pm 0.13$ M$_{\odot}$. Asssuming that the donor star 
is a 0.48 M$_{\odot}$ 
ZAMS, then $M_1\simeq1.6-1.9$ M$_{\odot}$ and $i\simeq 60-71^{\circ}$. On the
other hand, a canonical 1.4 M$_{\odot}$ would imply an evolved 
donor with $M_2\simeq0.35-0.42$ M$_{\odot}$ and $i\ge71^{\circ}$, which 
is difficult to reconcile with the absence of X-ray  eclipses, dips and the
low amplitude of the optical modulation (see Sect. 6.3). Therefore, a flaring 
angle of $\alpha\simeq12^{\circ}$ seems to support scenarios with ZAMS donors 
and massive neutron stars. These masses can be reproduced by recent 
evolutionary scenarios where some LMXBs are suposed to descend from binaries
with high mass donors (\citealt{schenker02}). These are significantly evolved 
at the begining of the mass transfer phase and mass is transferred in a thermal 
timescale until mass ratio reverses, which can result in the accretion of 
several tenths solar masses by the neutron star. This would make V801 Ara 
similar to other LMXBs such as 4U 1822-371 (\citealt{munoz05}). 

However, it is also possible to accommodate a 
canonical neutron star with a MS donor and a plausible inclination angle by 
pushing the disc flaring angle to higher values.   
For instance, $\alpha=16^{\circ}$ leads to $q\simeq0.30-0.34$,   
$f(M)\simeq 0.51$ M$_{\odot}$ and $M_1 \sin^3 i \simeq 
0.88$ M$_{\odot}$. Then, for $M_1=1.4$ M$_{\odot}$, 
$M_2\simeq0.45$ M$_{\odot}$ and $i\simeq57^{\circ}$. These masses would be 
consistent with standard evolutionary pictures where donors in LMXBs descend
from slightly evolved low-mass stars through dynamically stable mass transfer
and result in slightly undermassive stars.  
        
Note also that $\alpha<12^{\circ}$ are virtually ruled out 
because this would imply higher $K_2$ and $f(M)$ values which result in massive
neutron stars and very high inclinations. For instance, $\alpha=8^{\circ}$ 
leads to $f(M)\simeq 0.96$ M$_{\odot}$ and $M_1 \sin^3 i 
\simeq 1.5$ M$_{\odot}$. Then, for $i\leq76^{\circ}$ (lack of X-ray 
eclipses), $M_1\ge1.7$ M$_{\odot}$ 
whereas a plaussible $i\simeq57^{\circ}$ yields $M_1\simeq2.6$ M$_{\odot}$.   

The case of V926 Sco is unfortunately unconstrained because of the lack 
of an X-ray mass function. For example, assuming $\alpha=12^{\circ}$ leads to 
$q\simeq0.09--0.38$, $f(M)\simeq 0.22--0.68$ M$_{\odot}$ and $M_1 \sin^3 i 
\simeq 0.29--1.07$ M$_{\odot}$. The allowed range in $M_1$ and $M_2$ is too
wide as to impose any useful restriction on possible evoultionary scenarios.    
More, higher resolution data are required to measure the 
NIII $\lambda$4640 flux and $V_{\rm rot} \sin i$ as a function of orbital 
phase, from which tighter constraints on the inclination and disc flaring
angle can be set. This, 
together with the determination of the X-ray mass function for V926 Sco 
through pulse delays of burst oscillations and smaller errors in the 
$K_{\rm em}$ determinations, is expected to provide stronger  
limits on the stellar masses and the evolutionary models for these two LMXBs 
(see e.g. \citealt{munoz05}).
   
%------------------------------------------------------------------------------

%------------------------------------------------------------------------------

\section{Conclusions} 

The main results of the paper are summarized as follows:

\begin{itemize}

\item We have presented the first detection of the donor stars in the bursters 
LMXBs V801 Ara (=4U 1636-536) and V926 Sco (=4U 1735-444) through NIII
$\lambda$4640 fluorescent emission caused by irradiation. 

\item The narrow NIII $\lambda$4640 spots in the Doppler maps define 
$K_{\rm em}=277\pm22$ km s$^{-1}$ (V801 Ara) and $K_{\rm em}=226\pm22$ km
s$^{-1}$ (V926 Sco), and a new set of {\it spectroscopic} ephemerides which 
lend support to the assumption that photometric modulation is driven by the
visibility of the irradiated donor stars. On the other hand, the phasing of the
radial velocity curves of HeII $\lambda$4684 and the Bowen blend suggest
that these lines are mainly associated with the disc bulge. Our spectroscopic 
ephemerides, combined with burst oscillations data for V801 Ara, enables us 
to refine the neutron star's projected velocity to $K_1=90-113$ km s$^{-1}$ 
for this particular system.

\item Following \cite{munoz05}, we have computed the $K-corrections$ for 
the two LMXBs and obtain $K_2=360\pm74$ km s$^{-1}$, $q=0.21-0.34$, 
$f(M)=0.76\pm0.47$ M$_{\odot}$ (V801 Ara) and $K_2=298\pm83$ km s$^{-1}$, 
$q=0.05-0.41$, $f(M)=0.53\pm0.44$ M$_{\odot}$ (V926 Sco). 
Both systems are seen at intermediate inclination angles in the range 
$i\simeq30-60^{\circ}$, with V801 Ara having the higher inclination of the two.

\item Regarding V801 Ara, disc flaring angles $\alpha \le 8^{\circ}$ seem to be
ruled out because of the high inclinations implied. Opening angles 
$\alpha\simeq 12^{\circ}$ support massive neutron stars and main-sequence 
donors which may descend from intermediate-mass X-ray binaries as predicted by 
some evolutionary models (\citealt{schenker02}). Alternatively, higher opening 
angles $\alpha\simeq16^{\circ}$ are consistent with canonical neutron stars and   
main-sequence (or slightly evolved) donors which have evolved from standard
LMXBs through a dynamically stable mass tranfer phase. 

\item The lack of an X-ray mass function in V926 Sco prevents to set tighter 
constraints to the $K-correction$, mass estimates and evolutionary history in
this LMXB. 

\end{itemize}

\section*{Acknowledgments}

We thank the referee Thomas Augusteijn for helful comments to the manuscript. 
MOLLY and DOPPLER software developed by T.R. Marsh is gratefully 
acknowledged. JC acknowledges support from the Spanish Ministry of Science and
Technology through the project AYA2002-03570. DS acknowledges a Smithsonian 
Astrophysical Observatory Clay Fellowship as well as support through NASA GO 
grants NNG04GG96G and NNG04G014G. Based on data collected 
at the European Southern Observatory, Monte Paranal, Chile.

\bsp

\label{lastpage}

\end{document}